# The Unequal Incidence of Payroll Taxes with Imperfect Competition: Theory and Evidence

Felipe Lobel [*]

UC Berkeley

July 2022

**Abstract**

This paper provides a comprehensive examination of a Brazilian corporate tax reform targeted at the sector and product level. Difference-in-differences estimates instrumented by sector eligibility show that a 20 percentage point cut on payroll tax rates caused a 9% employment increase at the firm level, mostly driven by small firms. This expansion is not driven by formalization of existing workers, and it is explained by reduction on separations rather than additional hires. In terms of earnings, there is a significant 4% earnings increase in the long run, which is concentrated at leadership positions. The unequal pass-through worsen within-firm wage inequality. I exploit the exogenous variation on labor cost to document substantial labor market power in Brazil, where wages are marked down by 36%. Consistent with the empirical findings, I develop a model of factor demand with imperfect competition in the goods and labor market to shed light on the mechanism through which imperfect competition drives corporate tax incidence. The model is identified by the reduced form elasticities, and allows me to structurally estimate the capital-labor elasticity of substitution, which differs from the benchmark case of perfect competition.

[*]Department of Economics, lobel@berkeley.edu. This paper has previously circulated with the title "The Incidence of Payroll Taxation". I am indebted to my advisors Emmanuel Saez, Alan Auerbach, Patrick Kline, and Gabriel Zucman for their guidance and encouragement on this project. I would like to thank Nano Barahona, Michael Best, Bruno Barsanetti, Sydney Caldwell, Chris Campos, David Card, Carlos da Costa, Cláudio Ferraz, Fred Finan, François Gerard, Gustavo Gonzaga, Etienne Lehmann, Ted Miguel, Conrad Miller, Sebastián Otero, Jesse Rothstein, Jón Steinsson, João Thereze, Ricardo Perez-Truglia, Reed Walker, Chris Walters, Danny Yagan, and seminar participants at the NBER Business Taxation, NTA Annual Meeting, Zurich Conference on Public Economics, All Cal Labor Conference, IIPF Annual Congress, RIDGE Workshop on Public Economics, PacDev Conference, UC Berekeley Development, Public Finance and Labor Lunch for very helpful comments. The findings expressed in this paper are solely those of the author and do not represent the views of any other institutions.

# 1 Introduction

Payroll tax cuts are an expensive and pervasive[1] policy across the globe. On average, payroll taxes are responsible for 25% of total tax collection in OECD countries (OECD, 2019). These expensive policies are often rationalized by the classical assumption on aggregate labor demand being much more elastic than labor supply, which suggests that payroll taxes are borne by workers. Indeed the 2018 Congressional Budget Office relies on this assumption to predict the impact of payroll taxes in the US.

However, the community of scholars lack consensus on the labor market implications of payroll tax cuts. Part of the literature points out that workers bear the incidence, by showing zero employment, but positive pass-through to earnings (Gruber 1997; Gruber and Krueger 1991; Gruber 1994; Cruces, Galiani, and Kidyba 2010). At the other extreme, recent studies point to positive employment and zero earnings response (Saez, Schoefer, and Seim 2019; Kugler, Kugler, and Prada 2017). A third strand of literature reports results in between, with a partial pass through to earnings (Hamermesh 1979; Holmlund 1983; Kugler and Kugler 2009). At the center of this controversy there are two underlying questions: What are the labor market implications of a payroll tax cut? What is the role of firms' labor market power in determining the incidence of corporate taxes?

One reason for the lack of consensus in the literature is that most of the reforms studied in the past face at least one, out of the two common identification concerns. First, payroll tax cuts are typically targeted to specific workers (based on earnings, tenure or age), thus it is difficult to disentangle the effects of the reform from pay equity norms within firms. For example, if two workers perform similar tasks and differ across one dimension that is targeted by the policy, say worker's age, then it can be challenging for employers to differentiate wage of these similar workers (for a summary on the pay equity norms implications to labor market outcomes, see Dube, Giuliano, and Leonard 2019; Breza, Kaur, and Shamdasani 2018). Second, if the tax reform is implemented across the board, there might be other macro shocks able to confound the causal impact on the labor market.

In this paper, I overcome these challenges by exploiting a quasi-experimental large labor cost variation that targeted a small fraction of Brazilian firms[2], in the context of a payroll tax reform. The setting alleviates pay equity concerns, as all employees in a given firm face the same tax variation. At the same time, the Brazilian reform provided identification because not all firms were eligible for the tax cut. In December, 2011 the Government enacted a major corporate tax reduction aimed to reduce labor cost, in order to increase competitiveness of domestic firms. Initially, the policy targeted a few sectors and products, with gradual expansion of eligibility in subsequent years[3]. In the empirical design, I exploit this staggered implementation in a setting where most firms are never treated.

---

[1] US, Brazil, Chile, Italy, Colombia, Greece and Sweden are recent examples, just to cite a few.
[2] Less than 1.5% of total firms (or 110,000 firms) at the peak of its implementation.

At the firm level, the effect on average wages can be driven by pass through to wages and composition of the labor force. To disentangle these two underlying forces, I constructed two samples, one at the firm level and one at the worker level. I combine granular set of tax and labor administrative data on the universe of formal firms operating in Brazil between 2008 and 2017. To exploit regional variation on informality, I merge the data to the national Census. The final dataset provides a comprehensive laboratory of the Brazilian economy, and allows me to have a clear understanding of the responses to the tax reform.

I use this data to fit an event study model instrumented by sector eligibility to estimate the causal effect of the reform on the labor market. The importance of the IV in this context is because there is imperfect take-up in eligible sectors[4], and also because treatment is observed in non-eligible sectors due to the product eligibility criteria[5]. I show that being agnostic about these two margins of imperfect compliance lead to bias in the OLS estimates[6]. In fact, there are a series of papers studying the same reform that are not able to observe actual treatment due to lack of firm level tax data. They proxy the policy variation based on aggregated sector data[7]. The IV approach is only possible due to anonymized firm level tax data, which allows me to observe treatment at its most granular level. The data and econometric method allows me to conduct heterogeneity at the worker, firm and market level.

---

[3]IT, call center and lodging became eligible in 2012. Maintanance, transportation and media became eligible between 2013 and 2014

[4]Kleven and Waseem 2013 and Zwick 2021 provide evidence that some firms don't respond in tax dominated regions. This is consistent with what I observe in the tax data in Brazil even in the earlier years of the reform, when take-up was mandatory, but not enforced.

[5]Firm level tax data shows that treatment due to product eligibility is present in virtually every industry of the Brazilian economy.

[6]Note that these two sources of imperfect compliance lead to bias in opposite directions.

[7]Dallava 2014 finds null employment effects for most of the sectors, and positive employment effects in only a few subsectors of the IT industry. Scherer 2015 finds 15% employment increase. This study focuses in small firms and seeks identification based on the fact that the "Simples" tax regime (tax tier for some small firms) is not eligible for the reform. The main drawback of this approach is twofold: (i) there is substantial migration between "Simples" and the regular tax regimes; (ii) the treatment effect is measured on a subset of small firms that are more responsive than average, as I will show in a heterogeneity exercise. Baumgartner, Corbi, and Narita 2022 find a 5% employment increase. They assume perfect take-up rate on eligible sectors and restrict their sample to a few sectors that they claim to not be affected by the product eligibility criteria. It turns out that both of these assumptions don't match tax data. On top of more precision on the adjustments due imperfect compliance, my evaluation evaluation is also more comprehensive in two other aspects: (1) I consider the vast majority of economic sectors in Brazil, while previous work is restricted to a few specific sectors. (2) I analyze the reform from its beginning until recent years, while other studies were restricted to the three initial years of the program.

I find that the corporate tax cut causes a sharp expansion on firms' employment, with limited effects on earnings. The employment analysis is leveraged at the firm level to capture the effect of the tax reform on businesses. I find a 9% employment increase, which is mostly driven by small firms. This result is consistent with a broader literature that finds Government subsidies being more effective to boost employment on small business (Zwick and Mahon 2017; Criscuolo et al. 2019; Howell 2017; Bronzini and Iachini 2014). The setup of the Brazilian payroll tax reform is appropriate to connect with Industrial Policies because both of them offer shocks at the firm, rather than worker level.

Given the underlying payroll tax variation induced by the reform, the implied elasticity of employment with respect to labor cost is -0.71. The large employment effect doesn't affect the between occupation sorting of workers, and leads to a statistically significant positive effect on the average earnings at top percentiles of the within firm wage distribution. To analyze the pass through to earnings and minimize the contamination from workers' turn over, I follow the displacement literature (Jacobson, LaLonde, and Sullivan 1993, Lachowska, Mas, and Woodbury 2020) to build a sample of stable incumbent workers, who are assigned to treatment based on their pre-reform employers. At the worker level, I find an average 1.8% increase in earnings, which is sharp zero in the short run and a 4% significant effect in the long run. There is no significant difference across multiple workers characteristics, such as tenure, gender and race. However, it is largely heterogeneous on workers' occupation. While, workers in high skill occupations benefit from a significant 6% pass-through to earnings, low skill occupations experience a zero effect.

The identifying assumption is that conditional on fixed effects, eligibility is uncorrelated with time-varying unobserved determinants of employment and wage growth. I provide evidence that the identification assumption holds, and the estimates are robust to a wide variety of approaches. There are two main threats to identification. First, as in standard difference in differences, the design is compromised if parallel trends do not hold. This would be violated if the Government selects eligibility in a way that anticipates trends on the outcomes of interest. Second, the results would be biased if there were strategic selection into eligible sectors.

The formal and standard test for parallel trends is evaluating the statistical significance of pre-trends. I show not only that the pre-trends aren't statistically indistinguishable from zero in any of the outcomes, but also that eligibility is balanced in levels. Eligibility is not correlated with firms and workers characteristics in the pre-reform period. The result is robust to multiple estimation methods. As an alternative identifying strategy, I leverage a matching difference in differences to show that the results are qualitatively similar to the main empirical design. In this approach, I match each treated firm to a never treated that is similar in the pre-reform period. I also provide a more heuristic argument to highlight the arbitrary aspect of the eligibility criteria. Table 1, presents a non-exhaustive list of eligible versus non-eligible sectors that appear remarkably similar.

The second threat is about strategic selection into eligible sectors. I show that results are robust to eligibility assignment in the pre-reform period. Also, as a robustness check, I restrict to firms that have never changed sectors and the results are similar. I noticed from this exercise that very few firms actually change sectors, which suggests that this is not an easy margin of manipulation. When I focus on the firms that have changed sectors, I can show that there is not a trend of switching towards eligible sectors. All of these together is reassuring that the results are not driven by firms self selecting into eligible sectors.

Next, I turn the discussion to the mechanisms that rationalize the empirical findings. Informality is a natural candidate to consider in the context of developing economies (Ulyssea 2018; Haanwinckel and Soares 2021). One might suspect that the employment result is mechanically driven by formalization of existing informal workers, rather than an additional rise in employment caused by the reform. I exploit the Brazilian regional diversity in terms of informality to provide evidence that the employment expansion is not driven by highly informal areas. I also explore the transition from formal employment to non-employment/ informality to arrive at the same conclusion. The non-informality driven result can be rationalized by survey evidence[8] showing that most of the Brazilian informality is concentrated in self employment[9]. Along the same line, there is empirical evidence in Brazil and other developing countries that informal labor markets are segregated[10] (Alcaraz, Chiquiar, and Salcedo 2015; Dalberto and Cirino 2018), ie., formalization decisions go beyond a simple cost-benefit analysis outlined by the labor cost.

Taken together, the positive earnings effect as a response to labor cost reduction at the firm level, establishes evidence of labor market power. I develop a partial equilibrium model of factor demand with imperfect competition in the goods and labor market to interpret the findings. The model provides a framework to study corporate tax incidence in a imperfectly competitive economy. It follows the seminal ideas from Marshall 2009, formalized by Hicks 1932, and implemented by recent studies interested in understanding firm's adjustment decisions after inputs cost variation. I add value to the model, by adding imperfect competition.

---

[8]Pesquisa Nacional por Amostra de Domicílios (PNAD) is a household survey administered by the Brazilian Census Bureau (IBGE).

[9]Compared to informal employees, the incentives for formalization in the self-employment case is less dependent on payroll taxes, and more considerative of other variables such as: licenses to operate, costs relative to opening and maintaining a firm, other corporate taxes, legal liabilities, sanitary and security regulations.

[10]For a comprehensive analysis on the underlying forces of informality, refer to Perry 2007.

The model allows me to disentangle two forces driving the employment boost: the substitution from capital to labor, and the plant size expansion. In the presence of imperfect labor market competition, firms face higher labor costs as they expand, creating more pressure for inputs substitution.[11] For this reason, the structural estimate for the capital-labor elasticity of substitution identified by the reduced form elasticities is higher than in benchmark models, where labor markets operate in perfect competition. I show that in the limit case, where the elasticity of labor supply goes to infinity, the prediction of my model aligns with existent competitive models.

The rest of the paper is organized as follows. In Section 2 I discuss the institutional background and the data. Section 3 presents the empirical strategy and the main findings, including heterogeneity analysis. Section 4 discusses mechanisms. Section 5 develops the model and connects it to the data. Section 6 concludes.

## 2  Institutional Background and Data

The Brazilian payroll taxes are designed to fund social security programs, such as retirement pensions and unemployment insurance. In December 2011, the Government enacted a major corporate tax reduction aimed to reduce labor cost, and increase competitiveness for domestic targeted firms. The reform provides interesting variation because eligible and non-eligible firms present similar trends and levels in the period immediately before implementation.

### 2.1  Brazilian Payroll Tax System and the 2012 Reform

The Brazilian payroll tax system is similar to most OECD countries, however the tax reform was different. In Brazil, the reform was targeted at the firm level, while most of the reforms studied in the past were targeted at the worker level. This type of targeting provides an advantageous quasi-experimental design to study the labor market implications of payroll taxes on the labor market. Whereas, in worker level targeted reforms, the pass-through to wages can be confounded by pay equity norms.

---

[11]This intuition is clearly identified in the mathematical expressions.

The Brazilian payroll tax schedule has three components, and all of them are collected from firms. The main component is a 20% flat tax over the total wage bill. Secondly, there is an accident risk insurance component that varies between 1 to 3%[12]. The last layer of contribution is a 8 to 11% tax on wages, which is employee specific and can vary within workers of the same firm. All of these tax components are deposited in a social security fund that pools resources together. This implies that the public social security system does not provide individual savings accounts, where resources are traceable and mapped to specific workers' benefits.

On $14^{th}$ December, 2011 the Brazilian Federal Government announced the payroll tax cut program[13] that waived the main component of the payroll taxation, which means a tax cut equivalent to 20 percentage points of the total wage bill. To provide slight compensation to the Government budget in face of this large drop in tax collection, the benefited firms were imposed to pay a small 1 to 2.5% taxes on net of exports gross revenue. Figure 1 compares the payroll and revenue tax variation. In the figure, taxes are divided by the firm wage bill, and shows that the payroll tax drop is considerably larger than the revenue tax increase. This evidence showcases that the reform should be interpreted as a corporate tax cut, rather than a tax substitution.

Eligibility for the payroll tax exemption is sector and product specific. The first tax bill outlining the policies and the eligible sectors was passed in December 2011, and implemented a few months immediately after, April 2012. The reform was initially outlined on an executive bill that skipped prior Congress discussion. This type of corporate tax cut has never been implemented previously in Brazil, so this was not an expected policy by employers and employees. The policy is still valid nowadays[14], and there is no expectation of being eliminated in the near future.

---

[12]This tax varies according to the activity associated risk

[13]Law 12546/2011 approved by the Congress confirms Executive bill 540/2011 passed on August $2^{nd}$, 2011.

[14]As of September, 2022

Figure 1: Tax Implication of the Reform

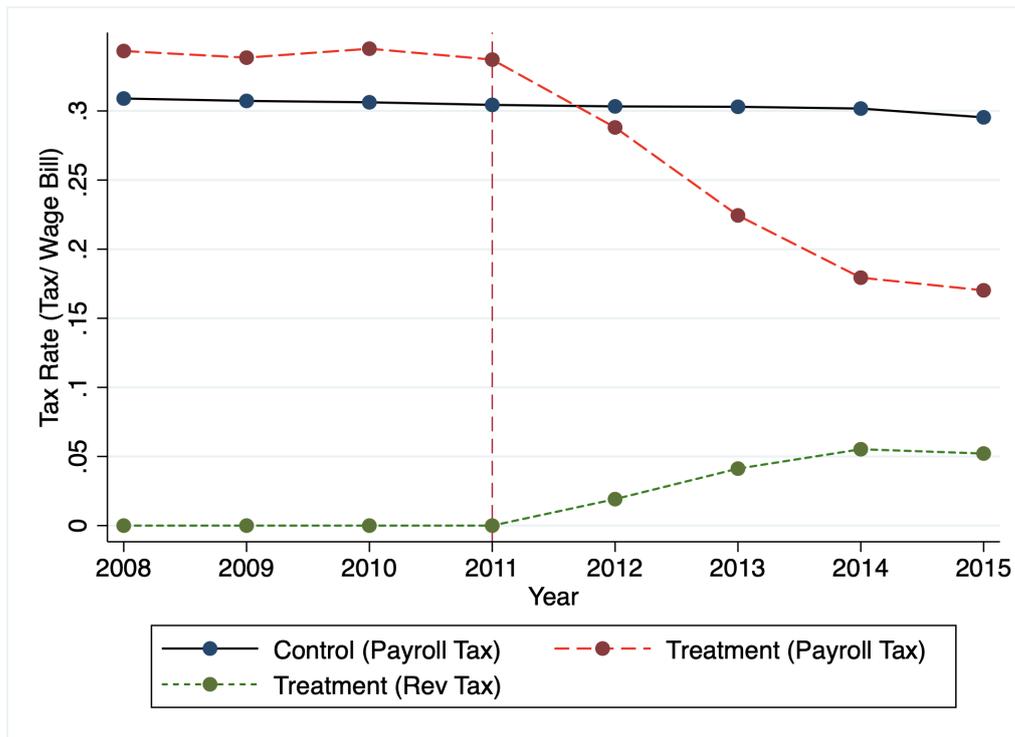

*Note:* This figure presents the evolution of tax rates for eventually treated vs control firms over the years. The blue line depicts payroll tax rates for control (never treated) firms, which slightly declined over the years, following global trends (OECD, 2019). The dashed red line represents the payroll tax rates for treated firms. The dashed green line presents the revenue tax rates that are substituted in once treatment takes place. Revenue tax rates are computed as a function of the total wage bill in order to facilitate comparisons.

The reform has a staggered implementation design. The first cohort of sectors became eligible in 2012. There were several other tax bills including more sectors to the reform in 2013 and 2014.[15] Another interesting variation is that within broad defined sectors, the reform did not provide eligibility to all subsectors. For example, when the media industry became eligible to the reform, open television were contemplated while cable television were not. Similarly, table 1 provides multiple examples of similar subsectors in broad defined industries, where one of them became eligible and the other not.

Regarding the product eligibility criteria, the tax bills define eligibility based on the Mercosur Common Nomenclature (NCM). Most of the product eligible firms are in the manufacturing industry, but treatment due to NCM criterion is not restricted to the manufacturing sector. Indeed, the vast majority of sectors in the Brazilian economy contain firms treated due to the product NCM criteria.[16] Treatment due to the NCM eligibility criterion only allows for partial payroll tax waive, according to the share of eligible products in the firms' gross income.

---
[15]IT, Call Center and Hotels were added in 2012. Retail, Construction and Maintenance were added in 2013. And a final wave in 2014 added Transportation, Infra-structure and Media sectors.

Over the years, 5 other tax bills[17] were passed promoting marginal changes to the program, such as modifying the revenue tax rates, or adding new sectors to the policy. One of the most relevant changes happened in December 2015 when the policy became less generous as the revenue tax rates increased from 1-2.5% to 1.5-4.5%[18]. At that moment, treatment assignment also became optional, which in practice is not a relevant change in the regime because even in the early years of the reform take-up in eligible sectors was far from perfect. Indeed, the imperfect take-up rate is a central aspect of the reform that deserves more discussion, as one might be puzzled to understand why an eligible firm wouldn't take such generous Government benefits.

There are a few facts that help to rationalize the imperfect take-up. First, the tax bills never mentioned any punishment to non-compliers. Possibly because from the legislative point of view eligibility was seen as beneficial to firms. Based on the Brazilian tax code it is implausible for prosecutors to suit firms that don't opt in a supposedly beneficial tax system. Second, enrollment in the program was not automatic as in the Swedish case studied by Saez, Schoefer, and Seim 2019.[19] In Brazil, firms have to self-report eligibility on Government provided software to enable tax exemptions[20], through separate tax forms. Figure 1 illustrates tax forms instructions and the set of information requested in the tax platform. Even though the tax substitution implied a net tax cut in most cases, empirical findings in other countries (Kleven and Waseem 2013 Zwick 2021) suggest that the operational filling process can lead to non responsiveness even in dominated tax regions.

The legislative decision process to define eligible sectors was political, and didn't seem to anticipate sector specific labor outcome levels, or trends. Section 3.3 is dedicated to provide details on the eligibility rules, and to test levels and trends of eligible sectors. The pre-trends observed in the event studies (figures 2 and 4) together with the pre-reform balance reported in table 2 is reassuring. Finally, the reform was not intended to increase deficits in the social security system. The

---

[16]This can be precisely observed in the anonymized micro tax data.

[17]Law 12546/2011, Law 12715/2012, Law 12844/2013, Law 13161/2015, Law 13202/2015, and Law 13670/2018.

[18]Law 13.161/2015

[19]In Sweden firms filled the same tax forms before and after the reform. Once firms provided information on their employees, the Tax Authority was the one computing firms' tax benefits.

[20]Firms inform eligibility on block 0 and this enables block P where tax relevant information is input.

Federal Treasury committed to cover any potential losses to the social security system. This is to say that the reform didn't affect individuals' perception on the solvency of their retirement plans.

## 2.2 Data and Descriptive Statistics

I constructed two samples, one at the firm and other at the worker level, by combining anonymized tax and labor administrative data on the universe of formal firms operating in Brazil between 2008 and 2017. To exploit regional variation on informality, I merge the data with the 2010 Census. The advantage of this data is that it allows me to track firms and workers over time, which constitutes an ideal laboratory to understand the effect of corporate tax policies on very granular measures of labor market outcomes.

**Labor Market Data.** For labor market data I use Relação Anual de Informações Sociais (RAIS), which is the matched employer-employee data set administered by the Ministry of the Economy. This data provides firm and worker level information covering every formal labor contract since 1976. I restrict the analysis to the period between 2008 and 2017, which allows me to track firms before and after the implementation of the payroll tax program. At the firm level, RAIS contains information on the tax regime[21], sector (at its most granular definition), total employment, wage bill, age and location. At the worker level, it contains variables regarding employment status, occupation, wage, race, gender, industry, municipality, as well as hiring and termination dates. Workers and firms are uniquely identified based on tax codes (PIS and CNPJ, respectively), which do not change over time. The main shortcoming in RAIS is the lack of information about informal and non-employed workers (see Dix-Carneiro 2014 for an overview of RAIS dataset).

I use other sources of administrative data to complement this dataset. The 2010 Census provides information that allows me to compute formalization rates at each of the 5,300 Brazilian municipalities. Finally, from the tax authority, I have firm level anonymized micro data on payroll tax, revenue tax, capital expenditure, and wage bill. Once this set of administrative data is merged, I construct two anonymized samples for the empirical analysis, one at the firm level, and the other at the worker level.

**Firm Level Sample.** To make the administrative data suitable to study the payroll tax reform in Brazil there are a few sample restrictions that are important to deal with specificities of the context. First, I exclude from the sample firms that have ever participated in the "Simples Nacional", which is a special tax tier not subjected to the payroll taxes studied in this paper. In the Brazilian corporate tax schedule there is a special tax tier named "Simples Nacional", which has never been subjected to the payroll taxes. The "Simples" is designed for small firms[22]. Firms in the "Simples" regime face a different tax tier which consolidates all tax liability in a single tax form with simplified and lower rates. Therefore, these firms are not eligible for the tax reform under analysis and neither are comparable to the firms in the regular tax tiers.

---

[21] There is a simplified tax regime ("Simples Nacional") targeted to small firms that are not subjected to the payroll tax cut under analysis.

In terms of sector comprehensiveness, the sample encompasses 19 out of 21 one-digit sectors[23] of the Brazilian economy. The construction sector is excluded because the treatment assignment to this sector was problematic. The tax bill allowed construction firms to be treated in only certain of its construction sites, according to the site's license date. This makes some of the construction corporations being partially treated, and, therefore, even with the firm level tax data it is not possible to observe the site level responses. Even if it was possible to observe the construction site level of granularity, this could be confounded by spillovers from non-treated sites within the same firm. Also, construction was at the epicenter of the "Car Wash" operation, a massive corruption scandal revealed in the decade of this study, which revealed that economic transactions on that sector were not responses to standard economic incentives of interest, but to illegal business negotiations.

The sector of repair and sale of motor vehicles was excluded to avoid lurking effects with other tax benefits conceived to these sectors in the period of analysis, in the context of Plano Brasil Maior. This was an industrial policy plan that focus on macro policies to improve economic conditions in the country. It equally affected control and treated firms, except for the aforementioned sector of repair and sale of motor vehicles, which receive specific tax exemptions, and for this reason we exclude from the analysis. Important to notice that these sectors are excluded at the one-digit (broadest) level, so it eliminates both treated and non-treated subsectors in these broad industries. The results are robust to alternative cleaning procedures such as winsorization and balanced panels. In the appendix, I repeat the analysis based on a winsorized data, in which wages and employment are winsorized at the 1 and 99% levels. In the second robustness check, I evaluate the results on a balanced panel of firms (the ones that appear in all ten years of the sample) to relieve concerns with firms' attrition.

---

[22]The current gross revenue eligibility threshold is BRL 4.8 millions (around USD 1 million).

[23]Sectors are defined according to Classificação Nacional de Atividades Econômicas (CNAE), which is administered by the National Statistics Bureau (Instituto Brasileiro de Geografia e Estatística).

**Worker Level Sample.** To maintain consistency between the firm and worker level analysis, I keep the same sample restrictions presented before to ensure an equivalent set of employers in both data sets. I follow the displacement literature (Jacobson, LaLonde, and Sullivan 1993; Lachowska, Mas, and Woodbury 2020; and Szerman 2019) to create a tenure restriction to track only workers that have been employed by the same employee for at least three years in the pre-reform period (2008-2011). This guarantees that results are driven by relatively stable employer-employee matches. This is important for two reasons. First, since we are interested on the incidence aspect, I want to minimize the turn over effect on wages, and focus on the pass-through interpretation. Second, since I assign workers to treatment based on the pre-reform period, it makes sense to consider workers with stable attachment to firms in order to classify them into treatment based on a meaningful employee-employer relationship. In the appendix, I show that removing the tenure constraint doesn't imply major changes to the results.

**Descriptive Statistics.** In the firm level sample there are 1,858,835 observations in the pre period (2008-2011). These firms are allocated in 19 one digit sectors that are broken down into 1,072 seven digit CNAE industries. Table 2 provides summary statistics for eligible and non-eligible firms in the pre-period (2008-2011). Prior to the tax reform, firm's average employment on December $31^{st}$ of each year was 55.34 workers. The average payroll tax rate was 31.78%, and 89% of employment was in low skilled occupations. On the workers level sample, table 3 reports average earnings of \$2,315.46 average earnings (approximately \$450 USD[24]). Workers average age is 39, the average share of male is 0.55, and average share of white is 0.67.

## 3 Main Findings

The corporate tax cut causes a sharp expansion on employment, with small but significant effects on long term wages. In this section, I present details about the main results, including heterogeneity analysis across firm size and workers characteristics.

### 3.1 Empirical Strategy

The main empirical strategy is an event study instrumented by the sector eligibility. The design explores the staggered implementation of the program, together with the fact that there is a large share of firms never eligible or treated. The IV is important to adjust for two margins: the imperfect take-up in eligible sectors; and the treatment in non-eligible sectors due to the product eligibility criteria. I fit similar models at the firm and worker level. Conditions for the LATE Theorem hold, thus IV estimates can be interpreted as average causal effects of tax cuts on compliers' outcomes. At the firm level the estimated structural equation is,

$$Y_{jt} = \sum_{k=-4, \neq -1}^{3} \beta_k D_{jt}^k + X'_{jt}\gamma + \alpha_j + \xi_{s1(j),t} + \epsilon_{jt} \qquad (1)$$

where, $Y_{jt}$ is the outcome of interest; $D_{jt}$ indicates that firm j is treated in year t; $X_{jt}$ are set of controls (e.g., education, gender, race, age and its square); $\xi_{s1(j),t}$ is 1-digit sector interacted with year fixed effect; $\alpha_j$ is the firm fixed effect; and k indexes the time relative to treatment.

---

[24] As of the exchange rate in September, $1^{st}$, 2022.

For each time t relative to treatment, there is one respective first stage equation. Thus, in total there are K first stage equations given by,

$$D_{jt}^k = \sum_{l=-4, \neq -1}^{3} \pi_{kl} \times \mathbb{I}(t = e_{s(j)} + l) \times L_{s(j)} + \alpha_j + \xi_{s1(j),t} + X'_{jt}\delta_k + \eta_{jt},$$

$$\forall k \in [-4, -2] \cup [0, 3] \quad (2)$$

where, $e_{s(j)}$ is the event date, in which firm j's sector becomes eligible; $L_{s(j)}$ indicates if firm j's sector is eventually eligible; and the remaining coefficients are the same as described before. Standard errors are conservatively clustered at the 5-digit industry-by-state level. Appendix C provides more details on the empirical model, and outlines the reduced form equations.

The event study design provides two main advantages. First, it validates the identifying assumption by showing that the pre-reform coefficients of interest are not statistically different from zero. Second, it provides intuition about the dynamics of the treatment effect relative to the year before the event. I combine the event study set up and the 2SLS framework to estimate the average treatment effect on compliers. The pooled version of the difference-in-differences model is outlined in equations 3 and 4.

$$D_{jt} = \pi L_{s(j)t} + \alpha_j + \gamma_t + \xi_{s1(j),t} + X_{jt} + u_{jt} \qquad (3)$$

where, $D_{jt}$ indicates that firm j is treated in year t; $L_{s(j)t}$ indicates that firm j belongs to a sector that is eligible for treatment and that period t is after the starting eligibility date for sector s(j); $X_{jt}$ are set of controls (e.g., education, gender, race, age and its square); $\xi_{s1,t}$ is 1-digit sector interacted with year fixed effect, $\alpha_j$ is the firm fixed effect. I cluster the standard errors at the level of the treatment variation (Bertrand, Duflo, and Mullainathan 2004 ; Cameron and Miller 2015). Because eligibility is defined at the industry level (mostly at the 7-digit industry level) and there is variation on state level tax on gross product, standard errors are conservatively clustered at the 5-digit industry-by-state level.

The first stage coefficient $\pi$ inflates as the take-up rate on treated sectors increases, and deflates as there are more treatments occurring in non-treated sectors due to the NCM criteria. The associated reduced form is expressed in equation 4,

$$Y_{jt} = \delta L_{s(j)t} + \alpha_j + \gamma_t + \xi_{s1(j),t} + X_{jt} + u_{jt} \qquad (4)$$

Identification relies on the assumption that conditional on fixed effects, eligibility is uncorrelated with time-varying unobserved determinants of employment and wage growth. This implies that in the absence of the reform, outcomes for eligible and non-eligible would follow similar trends. I test this assumption in a set of checks summarized in section 3.3. One of the tests consists in showing that the pre-reform coefficients of interest are not statistically significant.

The firm level sample can impose challenges to evaluate the earnings effect. One might be concerned that at the firm level, the average earnings can be affected by compositional changes in the labor force. To address this concern, I take advantage of the granularity of the micro data, to estimate a similar model at the worker level. I assign workers' eligibility status ({0,1}) according to their pre-reform employer, and then evaluate individuals' outcomes regardless of the firms that they end up working for. Thus, $L_{s(i,t_0)}$ is equal to one if firm j's pre-reform sector eventually becomes eligible. Similarly, to the firm level specification, the pooled difference-in-differences model at the worker level is given by,

$$D_{it} = \pi L_{s(j_0)t} + \theta_i + \alpha_j(i,t) + \xi_{s1(i,t_0),t} + X_{it} + u_{it} \qquad (5)$$

$$Y_{it} = \delta L_{s(j_0)t} + \theta_i + \alpha_j(i,t) + \xi_{s1(i,t_0),t} + X_{it} + u_{it} \qquad (6)$$

where, i indexes workers, $Y_{it}$ is workers' labor market outcome in year t; $\theta_i$ is the worker fixed effect; $\alpha_j(i,t)$ is the firm fixed effect; and the remaining variables and fixed effects are analogous to definitions in equations 3 and 4. Similarly to the firm level analysis, I also fit the event study model to the worker level sample. The structural and first stage equations are presented below,

$$Y_{it} = \sum_{k=-4, \neq -1}^{3} \beta_k D^k_{j(i,t_0)t} + X'_{it}\gamma + \theta_i + \alpha_{j(i,t)} + \xi_{s1(i,t_0),t} + \epsilon_{it} \quad (7)$$

$$D^k_{j(i,t_0)t} = \sum_{l=-4, \neq -1}^{3} \pi_{kl} \times \mathbb{I}(t = e_{s(i,t_0)} + l) \times L_{s(i,t_0)}$$
$$+ \theta_i + \alpha_{j(i,t)} + \xi_{s1(i,t_0),t} + X'_{it}\delta_k + \eta_{it}, \forall k \in [-4, -2] \cup [0, 3] \quad (8)$$

where, $D^k_{j(i,t_0)t} = 1$, if $t = e_{j(i,t_0)} + k$; $e_{j(i,t_0)}$ is the year when the pre-reform firm enters treatment; $e_{s(i,t_0)}$ is the year when the pre-reform sector becomes eligible; and the remaining variables and fixed effects are the same as defined before. Standard errors are conservatively clustered at the 5-digit industry-by-state level.

## 3.2 Results

In this section, I present the results on the employment boost at the firm level, and then the worker level effects on compensations. Workers benefit from a small, however positive and significant long run pass-through to compensation.

### 3.2.1 Firm-level

I fit equations 3 and 4 using the firm level data, to find a 9% employment increase, i.e., participation in the payroll tax program causes a 9% increase (SE = 0.0272) in the number of employees, for treated firms relative to control. Table 4 presents the estimates, which corresponds to an elasticity of employment with respect to labor cost of -0.71. The employment effect is driven by small firms, and it doesn't affect the between occupation sorting of workers (see column (3), table 4). There is a statistically significant effect on the average earnings at top percentiles of the within firm wage distribution. In terms of dynamics, figure 2 reports estimates from equations 1 and 2, which shows that as the reform kicks in, there is an immediate employment response that is sustained and slightly increased over time. The dashed horizontal line in the upper right part of the figure reports the local average treatment effect on compliers of 9% estimated based on equations 3 and 4.

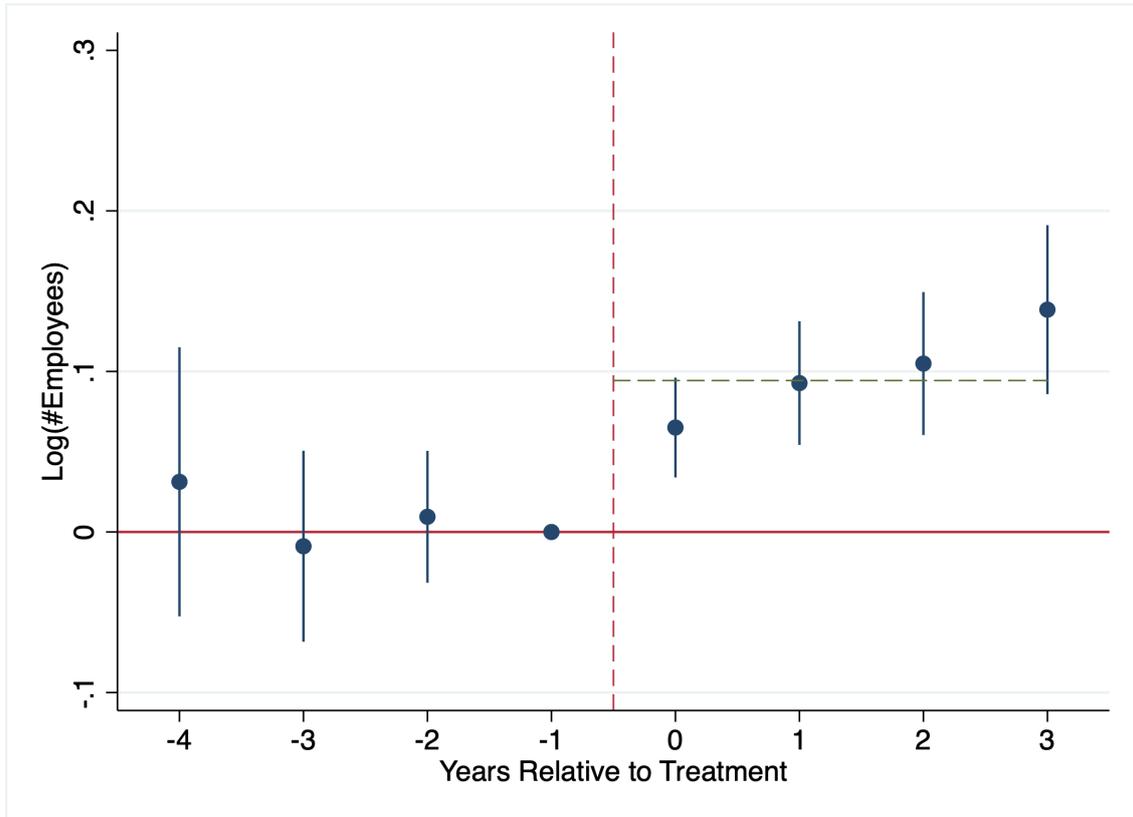

Figure 2: Employment: Event Study Estimates

*Note:* This figure presents the event study estimates for employment. The event is the year in which the firm enters treatment for the first time. I normalize the results with respect to one year prior to the event. The analysis spans three years prior to entering the payroll tax cut program and three years after. Standard errors are conservatively clustered at the 5-digit industry-by-state level.

Next, I fit equations 3 and 4 in three separate samples for small, medium and large firms. These categories are defined in the pre-reform period, i.e., prior to 2012. Firms are classified as small if they had less than nine employees, medium if they had between 10 and 49 workers, and large if they had more than 50 workers. Figure 3 reports the results on the size heterogeneity analysis. The blue markers show that the employment effect monotonically decreases with the firm size groups, and the employment increase is statistically different between small and large firms.

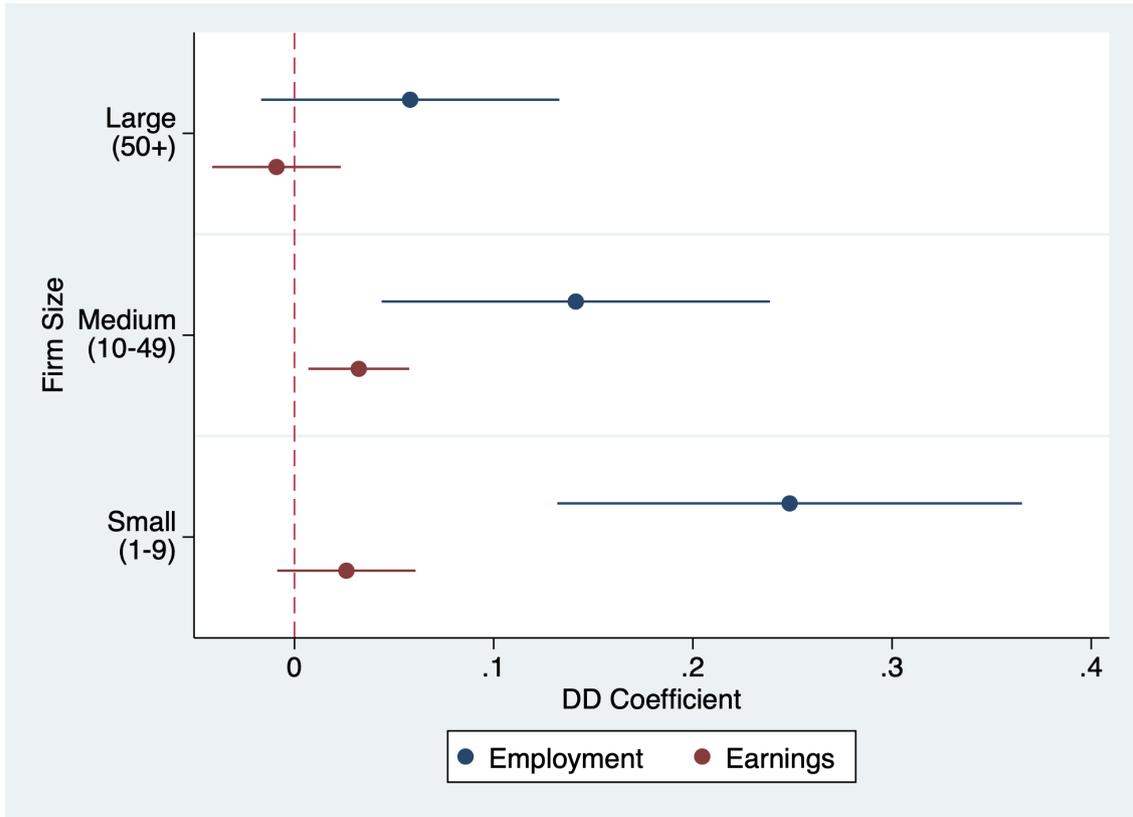

Figure 3: Firm Level: Heterogeneity Analysis

*Note:* This figure presents the event study estimates for the firm level estimates, for three firm size groups (small, medium and large firms). Size categories are defined in the pre-reform period. The blue marks plot the employment difference-in-differences coefficient, while the red markers plot the firm level earnings effect. Standard errors are conservatively clustered at the 5-digit industry-by-state level.

**Elasticities.** To compute the employment elasticity with respect to the labor cost, first we need to estimate the labor cost variation induced by the reform.[25] In the context of the Brazilian reform, the cost of labor is defined as the wage bill $\times$ (1 + payroll tax rate). Figure 9 plots firm level distribution of labor cost among treated and control firms, in the post period. Average labor cost for control firms is 131%, whereas for treated firms is 112%, which is consistent with the statutory rates. To estimate the labor cost variation, I rely on the IV outlined by equations 9 and 10. Equation 9 estimates the first stage, which adjusts for the imperfect compliance, and equation 10 is the reduced form, which estimates the labor cost variation on eligible firms.

---

[25] There is a revenue tax variation of small magnitude that I show in section 5 to be not relevant in interpreting the treatment effects.

$$D_{jt} = \alpha_j + \gamma_t + \xi_{s1(j),t} + \pi L_{s(j)t} + X_{jt} + u_{jt} \qquad (9)$$

$$\log(1 + \tau_{jt}) = \alpha_j + \gamma_t + \xi_{s1,t} + \delta L_{s(j)t} + X_{jt} + u_{jt} \qquad (10)$$

where, $\tau_{jt}$ is the payroll tax rate paid by firm j in year t; all other variables and fixed effects are identical to equations 3 and 4. As usual, the IV coefficient of interest is given by the ratio $\beta_{IV} = \frac{\delta}{\pi}$. Table 6 reports the tax cut impact on the labor cost. Column (1) shows that labor cost declines by 13.3% (SE = 0.002) according to the IV estimate, which aligns with the reform's statutory payroll tax cut (that varies from 131% to 112% of the total wage bill). It is reassuring that the IV estimates aligns with the statutory tax cut, as a sanity check to confirm that the IV is properly adjusting for the imperfect compliance. Column (2) reports a 6.75% (SE = 0.003) decline in the labor cost for eligible firms due to the tax reform. The impact on eligible firms is naturally smaller because some eligible firms don't face the payroll tax cut.

The elasticity of employment with respect to the labor cost (1 + payroll tax rate) is equal to -0.71[26]. In Saez, Schoefer, and Seim 2019, they find a smaller elasticity of employment with respect to labor cost (-0.21). However, there are three caveats in order to compare these results. First, in the Brazilian case since a smaller share of firms are treated it is expected to have more employment mobility across treatment status. Second, in the Swedish case they estimate the elasticity for young workers that can be different from the overall elasticity to all workers. One can imagine that the youth labor is cheaper and less price sensitive. Finally, in the Swedish tax reform there might be pay equity constraints limiting firms' ability to respond to the policy, thus implying lower elasticities.

The labor cost reduction generates an exogenous shift to labor demand, which allows me to estimate labor demand and labor supply elasticities with respect to compensation. I use the earnings response due to the exogenous variation on labor demand to compute the elasticities with respect to compensation.[27] I find $\epsilon_D = 0.57$, and $\epsilon_S = 2.78$. It is important to differentiate the market level labor supply elasticity to the firm level one. The former tends to be smaller[28] because when compensation is shocked at the firm level, the mobility across firms allow workers

---

[26] 9.44% divided by the payroll tax variation ($d\ln(1+\tau)$ = -0.133)

[27] The algebra for the computation of the elasticities are detailed in the appendix **??**

[28] Literature has estimated firm level labor supply around 2.5-4.5 and market level labor supply around 0.5.

to be more responsive compared to a market wide shock.

In the case of the Brazilian corporate tax reform, there was not a market level shock less than 2% of firms received the tax benefit. Eligible firms were distributed across multiple sectors with concentration on tiny defined industries that don't constitute a labor market themselves. There is considerable migration between eligible and non-eligible industries, which makes the labor supply interpretation much closer to the labor supply faced by the firm rather than market. In fact, in a recent study Lagos 2019 estimated that the labor supply elasticity faced by Brazilian firms is 2.88 (CI goes from 2.62 to 3.14). My estimate (2.78) lies in his 95% confidence interval. In a context of monopsony in the labor market, the labor supply elasticity corresponds to a mark down to wages of 36%.

**Within Firm Earnings Distribution.** To evaluate the distributional consequences of the earnings effect, I fit the event study models in equations 3 and 4 for a new set of outcome variables: average earnings per percentiles of the within firm distribution. Table 5 displays the aggregate estimates from equations 3 and 4. Column (1) reports the impact to the payroll tax waived firm's $99^{th}$ earnings percentile, represents the income of the top 1% workers in the organizations' hierarchy. They present a large and statistically significant increase of 4.2% (SE = 0.016), compared to the control. At the $90^{th}$ percentile (column 3), the payroll tax cut still created a large significant response of 2.12% (SE = 0.014) in the treated firms compared to the control. The effect shrinks as we move towards the bottom of the within firms earnings distribution, as displayed in columns (4) and (5). The distributional analysis is also implemented in an event study fashion to test if the parallel trend assumption holds at each percentile of the income distribution. In figures 10, 11, 12, 13 the pre-event coefficients are not statistically different than zero.

These results shed light to an important consequence of the tax policy, the within firm wage inequality. As the Government reduces payroll tax rates to lower labor cost, it increases the wage gap between high and low hierarchical levels. The discrepancy is even larger when considering the share of the wage bill paid to high versus low earnings workers. At the top of the distribution, wages were higher in the first place, and they are the ones receiving a higher percentage increase due to the tax reform.

**Occupation.** Next, I turn to study whether the firm level employment effect is driven by within or between occupations. One might wonder, if the firm expansion due to the corporate tax cut is driven by more employment of the same type of workers, or instead the firm employs from an upscale occupation position, to improve management over operational employees? To leverage this analysis I exploit the granularity of CBO occupation data, which contains 2,300 occupation codes. I ranked these occupations based on the pre-reform average earnings, and group them in percentiles according to the earnings ranking. Therefore, I can assign an index to each firm year based on the average occupation percentile that they employ from. Column (3) of table 4 shows that there is a sharp zero effect of the reform on firms' average occupation percentile. This fact favors the narrative that the tax reform expands employment within occupation, rather than between occupations.

### 3.2.2 Worker-level

To evaluate the pass-through of the tax benefit to incumbent workers, I fit equations 5 and 6 in the worker level sample. Even though, the gross earnings paid by the firm sharply drops after the reform (figure 6), I find that the net earnings (net of payroll taxes) received by employees presents only a modest increase of 1.8%, which is indistinguishable from zero at standard confidence levels. However, the results from equations 7 and 8 show that in the long run there is a 4% positive and statistically significant pass-through to net earnings. As depicted in figure 4, takes time for the earnings effect to show up and it only becomes significant three years after the tax cut.

Figure 4: Worker Level: Net Earnings Effect

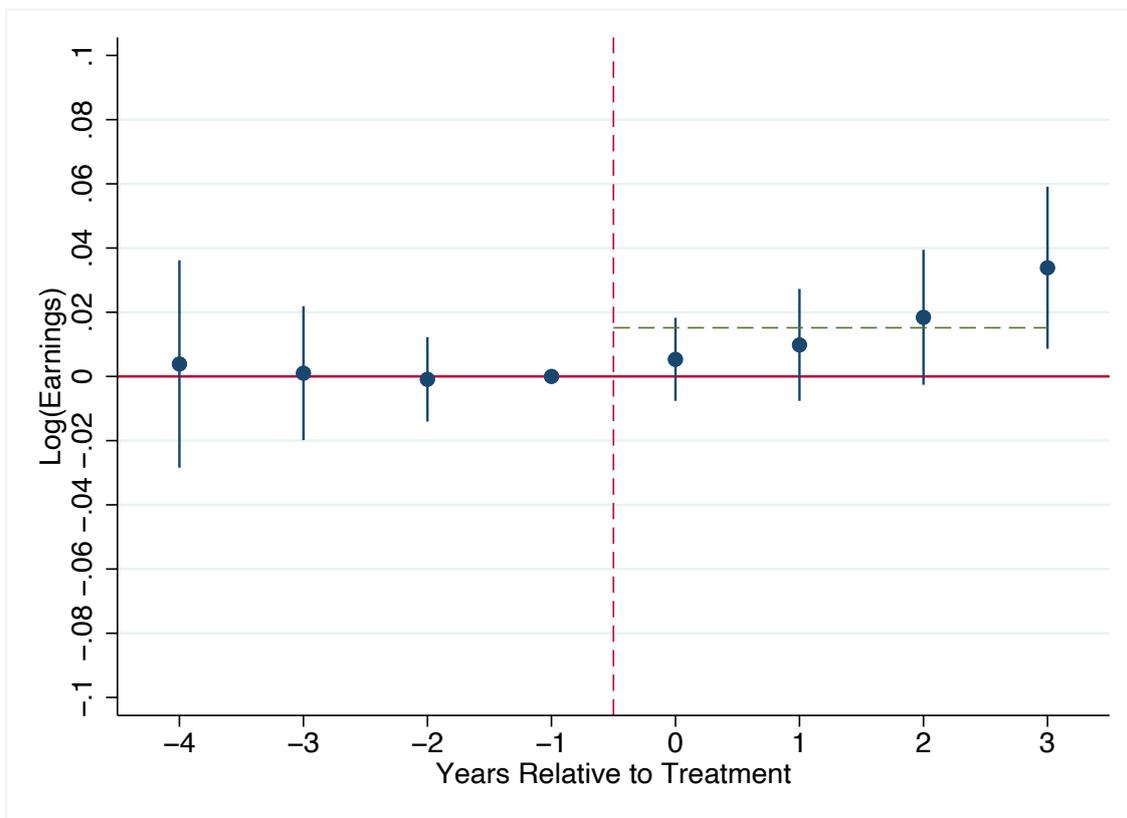

*Note:* This figure presents the event study estimates for average earnings (net of payroll taxes) for workers that were employed for at least three years in the same firm during the pre-reform period. I normalize the results with respect to one year prior to the treatment event. The analysis spans four years prior to the payroll tax cut program and three years after. The dashed horizontal line in the upper right part of the figure reports the local average treatment effect on compilers of 1.8% estimated based on equations 3 and 4. Standard errors are conservatively clustered at the 5-digit industry-by-state level.

**Heterogeneity.** I narrow the analysis on the earnings effect across many worker's characteristics, and I don't find relevant differences across most dimensions such as: unionization, pre-reform earnings and gender. I do find some differential pass-through based on race. The racial pay gap deteriorates after the tax benefit. There is a significant positive pass-through to white workers' earnings, while non-white employees face a sharp zero pass-through. Important to remember that this regression contains firm fixed effect, so the deterioration on the racial pay gap is not due to differential in the racial sorting into firms. The other margin of heterogeneity that has significant different pass-through effects is occupation, which I will explain next.

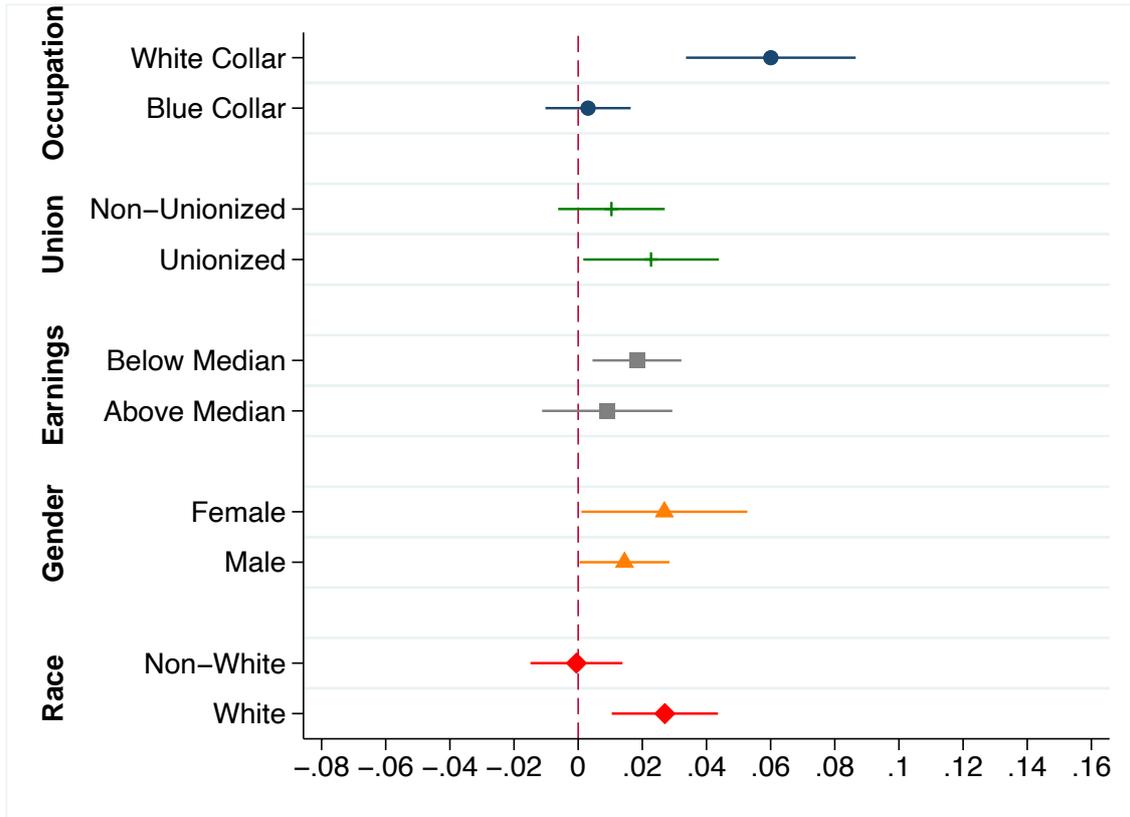

Figure 5: Worker Level: Summary of Heterogeneities on Earnings Effect

*Note:* This figure presents the pooled difference-in-differences coefficient for the earnings effect at the worker level, across many characteristics of interest, such as, income, tenure, gender and race.

**Occupation.** For this analysis, I rely on the CBO occupational code[29] to split employees into two occupation groups: leaders and operational workers. Leaders are directors, managers and qualified technical positions, while operational workers occupy the remaining positions. Figure 14 shows that there is a sharp pass-through to high skilled worker in the short run and grows over time. On the pooled diff-in-diff regression, the pass-through to high skilled workers is 6%, and sharp zero to low skilled workers. These findings are consistent with the within firm wage inequality presented in the previous section, and can be rationalized by larger labor supply elasticity to low skilled workers. Those are workers that offer less value added and their labor market operates closer to perfect competition, as a commodity type of labor.

**Gross Earnings.** In Brazil, firms are responsible to collect the payroll taxes, thus the difference between gross and net earnings is the gap between what employers pay and how much employees receive. Even though workers didn't observe substantial net earnings gains due to the reform (figure 4), it is important to note that firms did face a large and sharp decrease in the gross earnings paid to workers (figure 6). To compute the gross earnings, I use firm level annual tax and payroll data to obtain measures of firms' payroll tax rates per year. I apply these rates to workers net earnings to obtain the annual gross earnings of all workers in the sample. The pre-reform average gross earnings is $2,300 BRL and it drops $400 BRL (approximately 20p.p) immediately after the reform.

---

[29](Classificação Brasileira de Ocupação)

Figure 6: Worker Level: Gross Earnings Effect

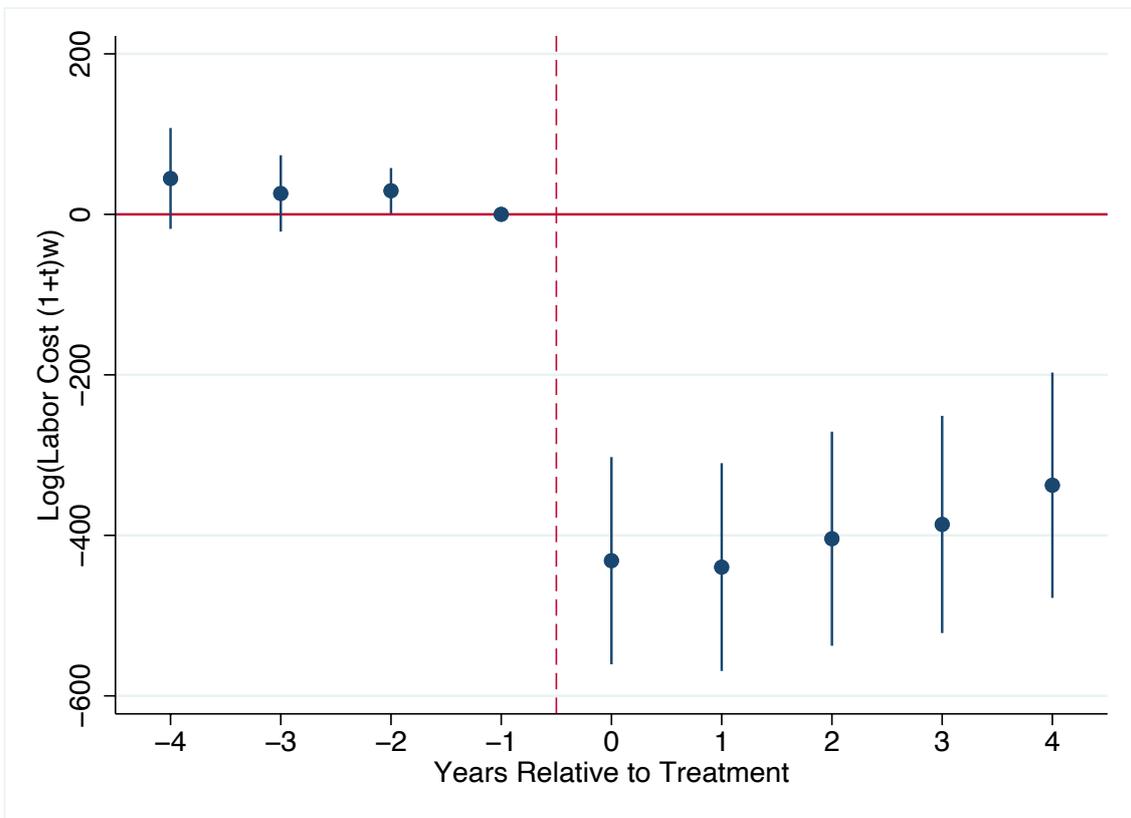

*Note:* This figure presents the event study estimates for average gross earnings paid workers that were employed for at least three years in the same firm during the pre-reform period. The labor cost is computed using firm level tax data, and worker level earnings data. I apply the firm payroll tax rate in year t, to all employees in that firm in year t. I normalize the results with respect to one year prior to the treatment event. The analysis spans four years prior to the payroll tax cut program and three years after. The plot shows an average decrease of $400 on the gross earnings, which has an approximate average of $2,300 during the pre-reform period. Standard errors are conservatively clustered at the 5-digit industry-by-state level.

**Minimum Wage.** One might wonder if the small earnings effect presented in figure

[4](#) is driven by the minimum wage constraint. The idea underlying this argument is that as labor demand expands, the new wage for minimum wage workers can still be under the minimum wage constraint. Thus for workers binding on the minimum wage, there won't be any observable earnings effect. Figures 15 and 16 suggest that this is not the case, as there is no statistical difference between the earnings effect for workers below and above the minimum wage barrier. To leverage this analysis, I classify workers into the minimum wage categories based on the modal pre-reform minimum wage status. On average 20% of workers in the sample are constrained by the minimum wage. More work is in progress to understand the reasons underlying the positive earnings effect for workers constrained by the minimum wage. One possibility is that the high informality levels, and the low minimum wage in Brazil makes the minimum wage likely to bind (or close to bind). This fact combined with the large shift in labor demand makes it reasonable to imagine that there is a positive average treatment effect even for those previously constrained by the minimum.

**Hire vs Layoff.** The large treatment effect on the number of employees per firm might raise the question of whether the employment boost is driven by increase in hires or decrease in separations. I conduct a set of analysis that show the separation channel as the driven force. I start by merely following the raw average of employment for treated and control firms, and notice that there is a decrease in the average employment of treated firms compared to control (see figure 17).[30] Second, I restrict the firm level sample to the set of stable incumbent workers, i.e., the ones employed pre-reform in both treated and control firms. Figure 18 shows that the employment local average treatment effect on this sample is qualitatively the same as in the main sample (with no restrictions), which suggests that the employment effect is due to the reduction on separations, rather than increase in hires.

### 3.3 Threats to Identification

The identifying assumption on the main difference-in-differences specification is that eligibility to the tax benefit is uncorrelated with the outcomes of interest, conditional on fixed effects. The main threat to the validity of this assumption is the Government anticipating sector specific trends when eligibility rules were defined. Another concern is that firms strategically select into sectors once the reform is announced. In this section, I provide multiple tests to address both of these concerns. I show that eligibility choices were a result of a political process that didn't anticipate sector specific trends. I also show that sector change is a difficult margin of manipulation and firms are not operating in this margin.

---

[30]The overall macro trend is due to a major recession experienced between 2014-16, in Brazil.

Regarding the concern with sector specific trends, I start by following the most standard and formal way to address this threat, which is testing the pre-trends. Second, on top of similar trends, I show that firms (and workers) are balanced in levels across eligibility groups. Third, I show that results are robust to alternative estimation methods, such as matching difference-in-differences. Fourth, I provide examples from the tax bill of non-eligible sectors remarkably similar to eligible ones. Regarding strategic selection into sectors, I first show that the results are robust to pre-reform sector assignment. Second, I show robustness to a sample that is restricted to firms that have never changed sectors. Third, I show that only a few firms have actually changed sectors, and there is not a trend of switching towards eligible sectors.

### 3.3.1 Selection on Eligibility

I rely on the event study design to show that the pre-reform coefficients of interest are not statistically significant, for any of the outcomes of interest. This means that treated and control groups were following similar trends when the reform was enacted. To address the balance in levels, I show in figures 19 and 20 that workers and firms' characteristics are not correlated with eligibility.[31] The baseline model shows that is unlikely that characteristics are able to explain eligibility choices. The one characteristic that is more concerning regarding balance is gender, which I control for in all specifications. Important to notice that the outcomes of interest are estimated in a two way fixed effect model. Relying on this model to test balance, I find sharp zero difference across eligibility groups (figures 19 and 20).

On top of the statistical test, I provide anecdotal evidence that the political process that determined eligibility was not seeking to anticipate sector specific trends. In table 1, I share a non-exhaustive list of similar sectors that are plausibly following the same trends, but present different eligibility status. For instance, the sector of trains were eligible, but touristic trains were not. The industry of open television is eligible, but cable television is not. The list goes on, and more examples can be found on table 1.

After all, if the reader is still not convinced that sector eligibility was not correlated with sector trends, I show that the results are qualitatively similar under alternative empirical strategies that rely on alternative identification assumption. I repeat the analysis using a matching difference-in-differences empirical strategy, in which I match each eventually treated firm to one never treated firm. Notice that the group of eligible sectors differ from treated firms because of the imperfect compliance discussed in section 2.1, thus the matching difference-in-differences strategy does not assume anything about the political process that defines eligibility. There are other threats to the matching difference-in-differences, that I test in a set of robustness checks described in the next paragraph. The bottom line is that the results are qualitatively similar in both strategies, which is reassuring.

---

[31] I do so by estimating the baseline OLS model: $L_{s(j)t} = X_{jt} + u_{jt}$, and the TWFE model: $L_{s(j)t} = X_{jt} + \alpha_j + \gamma_t + \xi_{s1(j),t} + u_{jt}$, where $Ls(j)t$ is a dummy to indicate if the firm (worker) is eligible in year t; $X_{jt}$ are characteristics of the firm (worker); and the fixed effects are the same used in all empirical specifications presented before.

The matching algorithm goes as follows. First, I match firms that belong to the same deciles on employment, wages and hires during the pre-reform years. A propensity score is fitted and applied to break eventual ties. The main concern with this approach is that firms can be similar in levels, but different in trends during the pre-reform period. I eliminate this concern by showing that the pre-trends are indistinguishable from zero. I ran a few other robustness tests in the matched sample. In one of them I assign placebo treatment at random and follow the same matching process to the placebo treated firms. As expected, the placebo tests generate zero employment and zero wage effects, providing evidence that the results are not driven by any inconsistency in the matching algorithm. In another test, I show that treated and control firms are balanced in levels of pre-reform characteristics.

### 3.3.2 Manipulation on Sectoral Choice

Another threat to the design is firms strategically changing sectors after the reform is announced. In that case, firms with expectations for employment growth could have self selected into treatment, and the results lose causal interpretation. I show in this section that sector manipulation is a difficult margin of evasion[32], and firms are not actively using it. I show in the data that there are only a small number of firms changing sectors, and even among those, there are not a trend of switching towards eligible sectors.

To reassure that change of sectors are not driving the results, I run a few extra robustness checks. First, I assign firms to eligibility based on their pre-reform sectors, and results remain qualitatively the same. Similarly, I restrict the sample to firms that have never changed sectors, and the results don't change. All these tests taken together, indicate that sector manipulation is not an active margin of response, which reinforces the causal interpretation of the results. This is not surprising as changing sectors is not operationally easy for firms. There is a long bureaucratic process for a firm to change sector. They need to obtain clearance from local tax authorities, civil registry offices and proof change of operational scope of its activities.

---

[32]Firms in the regular tax tiers (object of this study) face a long bureaucratic process to change sectors. They would first have to change their operating agreement, which requires proof that they are operating in a new industry. Then they need to request new operational licenses in multiple administration offices such as the city hall, state, federal tax authorities, and others. Failing in one of these steps can imply tax compliance fines.

# 4 Mechanisms

This section is dedicated to study mechanisms able to rationalize the employment and earnings effect of the payroll tax reform. First, I clarify through an extensive set of tests that results are not rationalized based on particular institutional settings of developing economies, such as informality. Finally, I provide evidence that the reduced form results are consistent with imperfect labor market competition.

## 4.1 Informality

I take advantage of the fact that Brazil is a large and diverse developing economy with some local labor markets that reassemble developed countries. I exploit this variation to disentangle the effects of a corporate tax reform in settings with different degrees of exposure to informality. As 45% of the Brazilian labor market is shadowed in the informal economy (PNAD, 2012)[33], one might be concerned that the employment increase due to payroll tax cut is mechanically driven by formalization of existing workers, rather than an additional rise in employment caused by the reform. I provide several evidences that this is not the case. First, I take advantage of the fact that two years prior to the payroll tax reform, the Brazilian Census Bureau implemented a national Census survey with rich regional informality data. There are 5,300 municipalities in Brazil which present a wide range of informality rates, see figure 21. At the lower end, municipalities present a formalization rate lower than 20% which are consistent with developing countries. At the upper tail, there are regions with more than 80% formalization rate which are standards of developed economies.

I split regions in two groups according to the position in the pre-reform median of the formalization rate. I leverage the analysis of labor market implications of the tax reform in both groups of regions. If the main employment response to the tax cut (figure 2) was driven by the mere formalization of informal workers, we should expect to see larger employment effects on high informality regions. I find precisely the opposite, i.e., low informality regions are the ones driving the employment effect, which is suggestive that the results are driven by additional employment rather than formalization of existing workers (figures 22 and 23).

---

[33]Pesquisa Nacional por Amostra de Domicílios (PNAD) is a household survey administered by the Brazilian Census Bureau (IBGE).

Since previous result suggested that small firms are driving the employment effect (figure 3), I also want to show that the firm size distribution is evenly distributed across pre-reform informality status, see figure 24. One might still be concerned that the labor cost variation induced by the policy in low and high informality areas can be different. I show in figure 25 that the first stage is uniform across informality status. I reinforce the informality analysis by tracking formal new hires that were previously formally employed. The matched employer-employee data allows me to track previous employment spells for workers that held formal jobs in the past. If the treatment effect were rationalized by hiring existent informal workers, we would see a sharp increase in the share of new hires coming from non-employment or informality. This is not what figure 26 suggests. The share of new hires coming from non-employment and informality is flat across time and across treatment status.

According to Ulyssea 2018, informal employment is concentrated in firms with lower average educational levels. In figure 27, I show that the employment effects are larger in firms with higher shares of qualified workers (less likely to have informal workers),[34] which adds as another evidence that the employment effect is not driven by informality. Finally, I show that the employment results holds, even if I restrict the sample to the set of incumbent workers. This result reinforces the idea that employment boost it is not driven by formalization of existing informal workers[35].

Two facts help to rationalize the informality findings. First is that informality in Brazil is mostly driven by self employment rather than informal employment (PNAD, 2012). There are reasons to believe that the self-employment formalization decision is less sensitive to payroll tax variation. Those workers are more comparable to entrepreneurs than employees. The formalization decision for self employed workers involves other costs such as: licenses to operate, costs relative to opening and maintaining a firm, other corporate taxes, legal liabilities, sanitary and security regulations. Second, even though there is a reduction on the labor cost, the worker's formalization decision goes beyond a simple cost-benefit analysis.[36] There are segregation between formal and informal labor markets (see Perry 2007

---

[34]I use the pre-reform educational level to split firms in below and above the median in the pre-reform share of workers with high school degree

[35]The sample of incumbents is based on formally employed workers in the pre-reform period.

[36]Alcaraz, Chiquiar, and Salcedo 2015; Dalberto and Cirino 2018

for discussion).

## 4.2 Imperfect Labor Market Competition

The Brazilian tax reform shifts treated firms' labor demand curve. In a perfectly competitive labor market, the only possibility for a positive pass-through to workers' compensation is a market level shock. This would be a scenario where the treatment makes worker's outside option more attractive, and the earnings respond to that even in the case of firms facing flat labor supply. I have already provided evidence that the reform does not generate market effects due to the policy design. Now, I formally test this result by computing the share of firms treated in each local labor market and analyzing the effect of the reform separately for firms in high versus low treated markets. If the positive earnings effect was driven by market effects, we should see higher pass-through for workers in highly treated labor markets. This is not what happened. Indeed, the payroll tax reform generates equal earnings effect across market level exposure to the policy. This reinforces the idea that the earnings effect in the Brazilian tax reform is driven by the positive labor supply elasticity faced by the firm, which constitutes clear evidence of imperfect labor market competition.

I follow Felix 2021 to define local labor market based on occupation x microregion cells. The firm level anonymized tax data allows me to compute the share of treated firms in each cell and separate workers and firms in above and below the median in terms of pre-reform market level treatment intensity. I separately estimate equations 7 and 8 and show in figures 29 and 30 that the pass-through is the same in highly versus mildly treated markets. In the next section, I take seriously the labor market friction into account, and incorporate the imperfect labor competition to advance on the understanding of payroll tax incidence.

## 5 Model and Structural Estimation

I interpret the empirical findings through the lenses of a Hicks-Marshall analysis. Recent papers have found this model useful to analyze the effect of minimum wage (Harasztosi and Lindner 2019), and capital tax subsidy (Curtis et al. 2021) on firms' production decisions. It is standard in this literature to consider imperfect competition in the output market (Hamermesh 1996). I extend this class of models to account for the empirically observed labor market power, which brings a few interesting insights. Monopsony power limits the employment responses to labor subsidies because the positive sloped labor supply curve faced by the firm imposes higher costs in case of employment increase. This force exacerbates the capital usage in the inputs composition, which will reflect on wages, profits, output quantity and prices.

To embrace all the elements observed in the Brazilian tax reform, I also extend the model to account for revenue taxes. I show that the effects of revenue taxes in the inputs mix and firm's outcomes depend on the number of firms subject to the tax reform and the tax rate. I analyze simultaneously all the tax variation imposed by the policy to bridge the model's predictions to the empirical findings. Relying on classical minimum distance method, I (will) simultaneously estimate the labor supply elasticities faced by the firms, the output demand elasticity and the capital-labor elasticity of substitution. This analysis provides evidence on how firms adjust to corporate tax reforms, and allows me to disentangle scale and substitution effects. Finally, the model allows me to understand the effects on prices, profits, and the overall incidence implications of the payroll tax reform. Notice that the profit effect is only possible to rationalize thanks to the monopolistic market structure.

## 5.1 Environment

Firms use capital and labor applied to a CES production function with constant returns to scale,

$$f = (s_L L^\rho + s_K K^\rho)^{\frac{1}{\rho}}$$

Firms face a positive sloped labor supply function characterized by the elasticity $\epsilon$ (Manning 2006; Card et al. 2018; Berger, Herkenhoff, and Mongey 2022). In the goods market, firms face competition a la Cournot and a constant elasticity of output demand $\eta$. The tax reform will lead adjustments along two margins: substitution and scale. First, in the profit optimization problem, firms decide the output level. For a chosen quantity, firms decide the inputs mix based on the cost minimization program. The complete derivation of the model can be found in appendix E.

## 5.2 Revenue Taxation

The effect of the revenue tax on revenue, capital and employment is,

$$\frac{\partial \log Rev}{\partial \log \tau} = \frac{\tau}{(1-\tau)} \frac{m}{m + (1-m)(1-\tau)} (1-\eta) = \nu(\eta) \qquad (11)$$

$$\frac{\partial \log K}{\partial \log \tau} = \frac{-\tau}{(1-\tau)} \frac{m}{m + (1-m)(1-\tau)} \eta = \xi(\eta) \quad (12)$$

$$\frac{\partial \log L}{\partial \log \tau} = -\chi(\epsilon, \eta, \rho) \frac{\tau}{(1-\tau)} \frac{m}{m + (1-m)(1-\tau)} \eta = \zeta(\epsilon, \eta, \rho) \quad (13)$$

where, m is the number of treated firms, $\eta$ is the output demand elasticity with respect to price, $\tau$ is the revenue tax rate faced by treated firms, and $\chi(\epsilon, \eta, \rho)$ is a combination of terms that arise in the solution (and are detailed in appendix E). In the absence of imperfect labor market competition the effect on capital and labor would be identical, as the revenue tax does not change the relative price of inputs.

Equations 11, 12 and 13 have a few intuitive interpretations. The revenue tax effect on revenue, depends mechanically on the tax rate, and on the share of firms subjected to the reform (due to price spillover). The elasticity $\eta$ makes the model versatile to accommodate policies targeted to a small set of firms versus market level interventions. If the shock is restricted to a reduced share of firms, which is the case in Brazil, one should expect higher value for $\eta$. In this case where demand is very price responsive, an increase in revenue tax implies a decrease in the firm's revenue, as equation 11 captures for $\eta$ greater than unit.

For the specific case of the Brazilian tax reform, the share of treated firms is small and the revenue tax rate is also small (around 1.5%). For both of these reasons, the effects depicted on equations 11, 12, and 13 is approximately zero. This result makes intuitive sense, as most of the action in this reform is on the payroll tax side, which I will evaluate next.

## 5.3 Payroll Taxation

What is the effects of payroll taxation in economies with imperfect labor market competition? In terms of employment, the result can be summarized in the following expression,

$$\epsilon_{L\theta} = \frac{\epsilon}{1 - \epsilon\rho + \epsilon} \left( \epsilon_{\lambda\theta} + \epsilon_{\lambda Q} \epsilon_{Q\theta} - 1 \right) + \left( \frac{(1-\rho)\epsilon}{1 - \epsilon\rho + \epsilon} \right) \epsilon_{Q\theta} \quad (14)$$

where, $\epsilon$ is the labor supply elasticity faced by the firm, $\rho$ is the technology parameter driving the capital-labor elasticity of substitution, $\theta$ is the labor cost given by (1 + payroll tax rate), and the remaining terms are relevant elasticities explained below.

Differently from the perfectly competitive case, the marginal cost is no longer a linear function of quantities (see proof of lemma 1, in appendix E). Since payroll taxes affect output level, the monopsony induced positive relationship between quantities and the marginal cost, creates an indirect effect of payroll taxes on the marginal cost. This relationship is represented by the interaction of the two elasticities ($\epsilon_{\lambda Q}\epsilon_{Q\theta}$) depicted in equation 14. The result is also affected by the direct effect of the payroll tax on the marginal cost $\epsilon_{\lambda\theta}$, and output $\epsilon_{Q\theta}$.

I derive closed form solution to the three colored elasticities ($\epsilon_{\lambda\theta}$, $\epsilon_{Q\theta}$, $\epsilon_{\lambda Q}$) in equation 14 that can be expressed as a function of observables and three parameters to be estimated ($\epsilon, \eta, \rho$). For presentation purpose, I summarize the result in the following expression,

$$\epsilon_{L\theta}(\epsilon, \eta, \rho) = \left(\frac{\epsilon}{1+\epsilon(1-\rho)}\right)\left(\Omega(\epsilon, \eta, \rho) - 1\right) \tag{15}$$

where, $\Omega(\epsilon, \eta, \rho)$ simplifies a combination of terms detailed in the Appendix. Analogously, the model allows me to compute the effect of payroll taxation on revenue and capital, as a function of observables and parameters to be estimated.

$$\epsilon_{K\theta} = \psi(\epsilon)\left[\left(\frac{1}{1-\rho}\right)\left(\frac{(\epsilon + 2\epsilon_{L\theta})(1-\eta)}{\epsilon + \epsilon_{L\theta}}\right)\right] \qquad \epsilon_{R\theta} = \psi(\epsilon)\left[(1-\eta)\left(\frac{\epsilon + 2\epsilon_{L\theta}}{\epsilon + \epsilon_{L\theta}}\right)\right]$$

where, $\psi(\epsilon)$ is a closed form term as a function of observables and the labor supply elasticity. The detailed derivation can be found in appendix E. In particular, if I take the limit of my model's estimate when $\epsilon$ goes to infinity,[37] I recover the exact same expressions derived in a standard Marshall-Hicks analysis and estimated by (Curtis et al. 2021) and (Harasztosi and Lindner 2019) for perfectly competitive labor markets. In this case, the substitution and scale effects are separable, as illustrated in the limits below.

$$\lim_{\epsilon \to \infty} \epsilon_{L\theta}(\eta, \rho) = \underbrace{\frac{-s_K}{1-\rho}}_{substitution} - \underbrace{s_L \eta}_{scale} \qquad \lim_{\epsilon \to \infty} \epsilon_{K\theta}(\eta, \rho) = \underbrace{\frac{s_L}{1-\rho}}_{substitution} - \underbrace{s_L \eta}_{scale}$$

### 5.4 Effects of the Brazilian Tax Reform

Putting the derivations from sections 5.2 and 5.3 together, I arrive at the model's prediction for firms' adjustments on labor, capital and revenue as a response to the

---

[37]This limit case moves the economy to a perfectly competitive labor market.

Brazilian corporate tax reform.

$$\beta_L = \frac{\partial \log L}{\partial Reform} = \overbrace{\epsilon_{L\theta}(\epsilon, \eta, \rho)\phi_1}^{\text{Effect from Payroll Tax}} - \underbrace{\xi(\eta)\phi_2}_{\text{Revenue Tax}} \quad (16)$$

$$\beta_K = \frac{\partial \log K}{\partial Reform} = \overbrace{\epsilon_{K\theta}(\epsilon, \eta, \rho)\phi_1}^{\text{Effect from Payroll Tax}} - \underbrace{\zeta(\epsilon, \eta, \rho)\phi_2}_{\text{Revenue Tax}} \quad (17)$$

$$\beta_R = \frac{\partial \log Rev}{\partial Reform} = \overbrace{\epsilon_{R\theta}(\epsilon, \eta, \rho)\phi_1}^{\text{Effect from Payroll Tax}} - \underbrace{\nu(\eta)\phi_2}_{\text{Revenue Tax}} \quad (18)$$

where, $\phi_1$ and $\phi_2$ measure the first stage associated with the policy, i.e., the percentage variation on tax rates induced by the reform. Mathematically, $\phi_1 = \frac{\partial \log w}{\partial Reform}$ and $\phi_2 = \frac{\partial \log \tau}{\partial Reform}$. Using anonymized tax data, I precisely estimate $\phi_1$ and $\phi_2$. As section 5.2 explains the effects from the revenue tax perturbation are muted in the Brazilian case. I have a perfect identified system of three equations (16, 17 and 18), and three parameters to be estimated $(\epsilon, \eta, \rho)$.

I (will) estimate the model using a classical minimum distance approach. This method consists in minimizing the sum of the squared difference between the elasticities computed in the model (equations 16, 17, 18) and data (quasi-experimental reduced form elasticities). This difference is weighted by the weighting matrix $\hat{W}$, which is equal to the inverse variance-covariance matrix of the empirical moments. Formally, the program minimizes, $[\hat{\beta} - m]'\hat{W}[\hat{\beta} - m]$, where $m(\epsilon, \eta, \rho) = [\epsilon_{L\theta}, \epsilon_{K\theta}, \epsilon_{R\theta}]$, and $\hat{\beta} = [\hat{\beta}_L, \hat{\beta}_K, \hat{\beta}_R]'$.

The complete structural estimation is ongoing as the reduced form estimation for equations 17, and 18 is in progress. However, I do have a robust estimate for equation 16. For this final exercise, I assume reasonable values for the parameters $(\epsilon, \eta)$ based on a wide range of estimates from previous work, and use equation 16 to estimate the capital-labor elasticity of substitution $(\sigma = \frac{1}{1-\rho})$.

Moving from the top green line to the bottom black line in figure 7, I am increasing the labor supply elasticity. As $\epsilon \to \infty$, the estimates for $\sigma_{KL}$ in my model converges to the perfectly competitive case. This shouldn't be a surprise since I have already shown that my model collapses to the perfect competitive in the limit case. The main point of the exercise in figure 7 is to show that the estimates for the elasticity of substitution between capital and labor is sensitive to the market structure assumed for the labor market. In the presence of monopsony, not accounting for this market friction, can generate a bias (from 28 to more than

Figure 7: Elasticity of Substitution for Wide Range of $(\epsilon, \eta)$

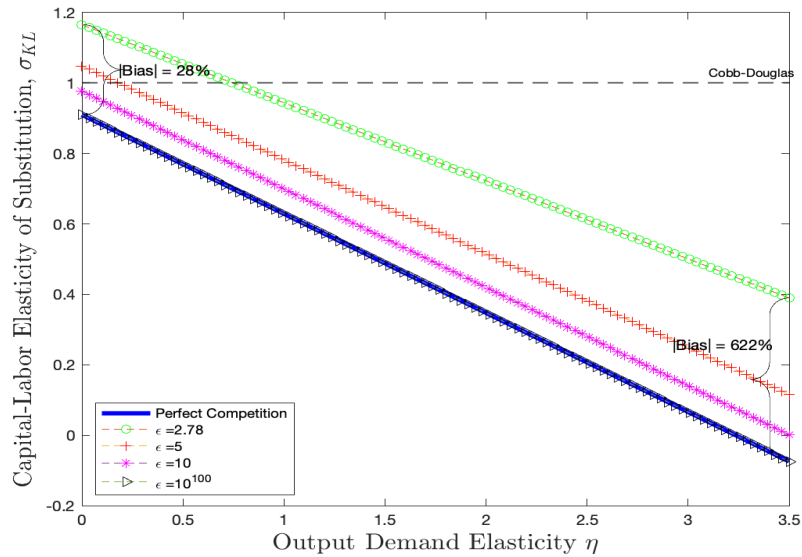

*Note:* This figure presents the sensitivity of $\sigma_{KL}$ with respect to $\eta$ and $\epsilon$. The blue solid line shows the estimates for a model under perfect competition. The other lines present the estimates for a wide range of labor supply elasticities. The range of $\eta$ varies from 0.11 used by Harasztosi and Lindner 2019 and 3.5 used by Curtis et al. 2021.

600%[38]) on the estimates for the capital-labor elasticity of substitution.

# 6 Conclusion

In this paper I show that despite of being an expensive policy, payroll tax cuts are effective to extend employment spells. However, they come at the cost of worsening within firm wage inequality. I show the role of imperfect labor market competition in rationalizing the incidence of payroll taxes and the firms' margins of adjustment. For example, monopsony power amplifies incentives to substitute capital to labor after a payroll tax reduction, which has direct implications on firms' ability to appropriate from the tax subsidies.

---

[38]This range depends on the output demand elasticity as depicted in figure 7.

# References

Alcaraz, Carlo, Daniel Chiquiar, and Alejandrina Salcedo. 2015. *Informality and segmentation in the Mexican labor market.* Technical report. Working Papers.

Baumgartner, Erick, Raphael Corbi, and Renata Narita. 2022. *Payroll Tax, Employment and Labor Market Concentration.* Technical report. University of Sao Paulo (FEA-USP).

Berger, David, Kyle Herkenhoff, and Simon Mongey. 2022. "Labor market power." *American Economic Review* 112 (4): 1147–93.

Bertrand, Marianne, Esther Duflo, and Sendhil Mullainathan. 2004. "How much should we trust differences-in-differences estimates?" *The Quarterly journal of economics* 119 (1): 249–275.

Breza, Emily, Supreet Kaur, and Yogita Shamdasani. 2018. "The morale effects of pay inequality." *The Quarterly Journal of Economics* 133 (2): 611–663.

Bronzini, Raffaello, and Eleonora Iachini. 2014. "Are incentives for R&D effective? Evidence from a regression discontinuity approach." *American Economic Journal: Economic Policy* 6 (4): 100–134.

Cameron, A Colin, and Douglas L Miller. 2015. "A practitioner's guide to cluster-robust inference." *Journal of human resources* 50 (2): 317–372.

Card, David, Ana Rute Cardoso, Joerg Heining, and Patrick Kline. 2018. "Firms and labor market inequality: Evidence and some theory." *Journal of Labor Economics* 36 (S1): S13–S70.

Criscuolo, Chiara, Ralf Martin, Henry G Overman, and John Van Reenen. 2019. "Some causal effects of an industrial policy." *American Economic Review* 109 (1): 48–85.

Cruces, Guillermo, Sebastian Galiani, and Susana Kidyba. 2010. "Payroll taxes, wages and employment: Identification through policy changes." *Labour economics* 17 (4): 743–749.

Curtis, E Mark, Daniel G Garrett, Eric C Ohrn, Kevin A Roberts, and Juan Carlos Suárez Serrato. 2021. *Capital investment and labor demand.* Technical report. National Bureau of Economic Research.

Dalberto, Cassiano Ricardo, and Jader Fernandes Cirino. 2018. "Informalidade e segmentacao no mercado de trabalho brasileiro: evidencias quantilicas sob alocacao endogena." *Nova Economia* 28:417–460.

Dallava, Caroline Caparroz. 2014. "Impactos da desoneracaoo da folha de pagamantos sobre o nivel de emprego no mercado de trabalho brasileiro: Um estudo a partir dos dados da RAIS." PhD diss.

Dix-Carneiro, Rafael. 2014. "Trade liberalization and labor market dynamics." *Econometrica* 82 (3): 825–885.

Dube, Arindrajit, Laura Giuliano, and Jonathan Leonard. 2019. "Fairness and frictions: The impact of unequal raises on quit behavior." *American Economic Review* 109 (2): 620–63.

Felix, Mayara. 2021. "Trade, Labor Market Concentration, and Wages." *Job Market Paper.*

Gruber, Jonathan. 1994. "The incidence of mandated maternity benefits." *The American economic review:* 622–641.

———. 1997. "The incidence of payroll taxation: evidence from Chile." *Journal of labor economics* 15 (S3): S72–S101.

Gruber, Jonathan, and Alan B Krueger. 1991. "The incidence of mandated employer-provided insurance: Lessons from workers' compensation insurance." *Tax policy and the economy* 5:111–143.

Haanwinckel, Daniel, and Rodrigo R Soares. 2021. "Workforce Composition, Productivity, and Labour Regulations in a Compensating Differentials Theory of Informality." *The Review of Economic Studies* 88 (6): 2970–3010.

Hamermesh, Daniel S. 1979. "New estimates of the incidence of the payroll tax." *Southern Economic Journal:* 1208–1219.

———. 1996. *Labor demand.* princeton University press.

Harasztosi, Péter, and Attila Lindner. 2019. "Who Pays for the minimum Wage?" *American Economic Review* 109 (8): 2693–2727.

Hicks, John R. 1932. "Marginal productivity and the principle of variation." *Economica,* no. 35: 79–88.

Holmlund, Bertil. 1983. "Payroll taxes and wage inflation: The Swedish experience." *The Scandinavian Journal of Economics:* 1–15.

Howell, Sabrina T. 2017. "Financing innovation: Evidence from R&D grants." *American Economic Review* 107 (4): 1136–64.

Jacobson, Louis S, Robert J LaLonde, and Daniel G Sullivan. 1993. "Earnings losses of displaced workers." *The American economic review:* 685–709.

Kleven, Henrik J, and Mazhar Waseem. 2013. "Using notches to uncover optimization frictions and structural elasticities: Theory and evidence from Pakistan." *The Quarterly Journal of Economics* 128 (2): 669–723.

Kugler, Adriana, and Maurice Kugler. 2009. "Labor market effects of payroll taxes in developing countries: Evidence from Colombia." *Economic development and cultural change* 57 (2): 335–358.
35

Kugler, Adriana, Maurice Kugler, and Luis Omar Herrera Prada. 2017. *Do payroll tax breaks stimulate formality? Evidence from Colombia's reform.* Technical report. National Bureau of Economic Research.

Lachowska, Marta, Alexandre Mas, and Stephen A Woodbury. 2020. "Sources of displaced workers' long-term earnings losses." *American Economic Review* 110 (10): 3231–66.

Lagos, Lorenzo. 2019. "Labor Market Institutions and the Composition of Firm Compensation: Evidence from Brazilian Collective Bargaining." *Job Market Paper.*

Manning, Alan. 2006. "A generalised model of monopsony." *The Economic Journal* 116 (508): 84–100.

Marshall, Alfred. 2009. *Principles of economics: unabridged eighth edition.* Cosimo, Inc.

Perry, Guillermo. 2007. *Informality: Exit and exclusion.* World Bank Publications.

Saez, Emmanuel, Benjamin Schoefer, and David Seim. 2019. "Payroll taxes, firm behavior, and rent sharing: Evidence from a young workers' tax cut in Sweden." *American Economic Review* 109 (5): 1717–63.

Scherer, Clovis. 2015. "Payroll tax reduction in Brazil: Effects on employment and wages." *ISS Working Paper Series/General Series* 602 (602): 1–64.

Szerman, Christiane. 2019. "The employee costs of corporate debarment." *Available at SSRN 3488424.*

Ulyssea, Gabriel. 2018. "Firms, Informality, and Development: Theory and Evidence from Brazil." *American Economic Review* forthcoming.

Zwick, Eric. 2021. "The costs of corporate tax complexity." *American Economic Journal: Economic Policy* 13 (2): 467–500.

Zwick, Eric, and James Mahon. 2017. "Tax policy and heterogeneous investment behavior." *American Economic Review* 107 (1): 217–48.

# Appendix

Figure, tables, mathematical proofs and robustness are in the online appendix.